# GranularGPT: Natural-Language-Driven End-to-End Discrete Element Method Simulations

Erlei Li, Sirui Tang, Ai Wang, Mingwei Hu, Yuchen Dai, Geoff Wang*

*School of Chemical Engineering, The University of Queensland, St. Lucia, 4072, Australia*

* *Corresponding author: gxwang@uq.edu.au (Geoff Wang)*

**Abstract**: Discrete element method (DEM) simulations are widely used to investigate granular and particulate systems, but their application often requires specialized expertise in particle physics, numerical modelling, and simulation post-processing. Large language models (LLMs) offer new opportunities to simplify these workflows by enabling users to interact with simulation tools through natural language. This study offers GranularGPT, an LLM-assisted agentic framework for end-to-end DEM simulation workflows, from natural-language problem descriptions to simulation execution and automated post-processing. The framework enables users to generate DEM simulation cases, execute and monitor simulations, and automatically extract and report physical and process-level metrics. Evaluation across 60 normal-condition cases covering 12 DEM scenarios and parameter variations achieve a simulation success rate of 91.7%. In the error-handling evaluation, all 10 repairable errors are detected and 9 are successfully resolved, while all 5 non-repairable cases are correctly identified and reported. The framework automatically extracts a range of DEM-relevant metrics, including particle size distributions, packing and density measures, velocity profiles, discharge rates, mixing indices, contact force characteristics, and thermal quantities. GranularGPT provides a foundation for improving the accessibility and efficiency of DEM simulation workflows while reducing the specialized expertise required for their operation.

## 1. Introduction

Powder and particle technologies underpin a broad range of engineering and manufacturing processes, including granular transport [1-3], pharmaceutical processing [4-6], mining [7, 8], agricultural handling [9, 10], additive manufacturing [11-13], and chemical production [14, 15]. In these systems, macroscopic behaviour emerges from the collective dynamics of large numbers of discrete particles. Interparticle and particle-boundary interactions, including contact, friction, rolling resistance, cohesion, and gravity-driven rearrangement, can strongly influence the resulting bulk behaviour. Depending on particle properties, loading conditions, geometry, and external forcing, granular materials can exhibit solid-like, fluid-like, or gas-like behaviour, making their prediction challenging. Computational modelling therefore plays an important role in understanding particle-scale mechanisms, interpreting granular behaviour, and supporting the design and optimization of particle-based processes. Discrete element method (DEM) simulations [16, 17] are particularly valuable because they provide access to particle-scale information that is difficult to obtain experimentally or represent directly using continuum models.

The DEM has become an established numerical approach for modelling granular and particulate systems [18, 19]. Originally introduced for assemblies of discrete particles, DEM explicitly tracks particle motion and evaluates interparticle and particle-boundary interactions during the simulation. This particle-scale description enables investigation of phenomena such as packing [20, 21], segregation [22-24], mixing [25-27], flowability [28, 29], and discharge behaviour [30-32]. Open-source and commercial DEM platforms have subsequently expanded the applicability of the method across academic and industrial problems. Particularly, LIGGGHTS provides an open-source DEM environment and has been widely used for granular and particulate simulations [33, 34]. However, the increasing sophistication of DEM applications has also increased the complexity of the associated simulation workflows. The practical challenge is often not simply solving the governing equations, but constructing and managing a complete simulation pipeline involving case setup, parameter specification, execution, monitoring, restart, parameter variation, output generation, and post-processing.

Developing such a workflow requires substantial domain and software-specific expertise. A user must determine process physics, appropriate particle and material properties, contact models, particle-generation procedures, geometries, boundary conditions, numerical parameters, and output quantities, while also translate these choices into the syntax and capabilities of a particular DEM solver. Post-processing introduces another layer of complexity because simulation outputs often need to be converted into physically meaningful quantities through additional scripts or analysis procedures. Consequently, even when the underlying physical problem is well understood, moving from a conceptual description of a granular system to a complete, executable, and analysable DEM case can require considerable manual effort and specialized expertise.

Recent advances in large language models (LLMs) have created new possibilities for interacting with computational tools through natural language. LLMs have demonstrated strong capabilities in code generation, instruction following, technical information processing, and interactive problem solving [35, 36], providing a potential interface between high-level user intent and software execution. More recently, LLM-based scientific agents have extended this paradigm by combining language models with external tools, domain knowledge, iterative reasoning, and computational feedback to perform multi-stage scientific tasks [37-39]. These systems have been explored for scientific workflows, molecular modelling [40-42], computational fluid dynamics [43, 44], and other computational research tasks [45-47]. However, an end-to-end LLM-assisted workflow specifically designed for DEM, covering the transition from natural-language problem descriptions through DEM execution and quantitative post-processing, remains comparatively underexplored.

In this work, we present GranularGPT, an LLM-assisted agentic framework for end-to-end DEM simulation workflows. The framework translates natural-language descriptions into executable LIGGGHTS simulations and coordinates simulation setup, execution, monitoring, automated post-processing, and quantitative reporting. Structured case representations, domain knowledge, physics-aware validation, and error repair are incorporated to improve workflow reliability. We evaluate GranularGPT across multiple DEM scenarios, parameter variations, and error-handling cases, assessing its ability to construct and execute simulations, recover from selected errors, and extract quantitative results. The results demonstrate the feasibility of using LLMs to support DEM workflows from natural-language problem descriptions to executable simulations and automated analysis.

## 2. Method

### 2.1 Design philosophy and system architecture

This work presents GranularGPT, an LLM-assisted agentic framework for end-to-end DEM simulation and analysis in LIGGGHTS. The framework enables users to describe a DEM problem in natural language and coordinates the subsequent workflow from simulation-case construction and solver execution to monitoring, automated post-processing, and quantitative reporting. The objective is not simply to generate a LIGGGHTS input script, but to establish a unified workflow that connects natural-language problem descriptions with executable simulations and interpretable quantitative results.

As illustrated in Fig. 1, GranularGPT comprises six functional components: (1) natural-language interpretation, (2) knowledge-guided DEM representation, (3) structured case-schema construction, (4) physics-aware validation and conservative repair, (5) deterministic simulation generation, and (6) simulation execution and automated analysis. Starting from a natural-language request, the framework identifies the required physical, numerical, geometric, and workflow information and maps these requirements to a structured DEM case representation. The case is then enriched with solver-specific information and evaluated for physical and numerical consistency before being converted into an executable LIGGGHTS simulation. Following execution, the framework manages runtime information and simulation outputs and automatically performs the requested post-processing and quantitative analysis.

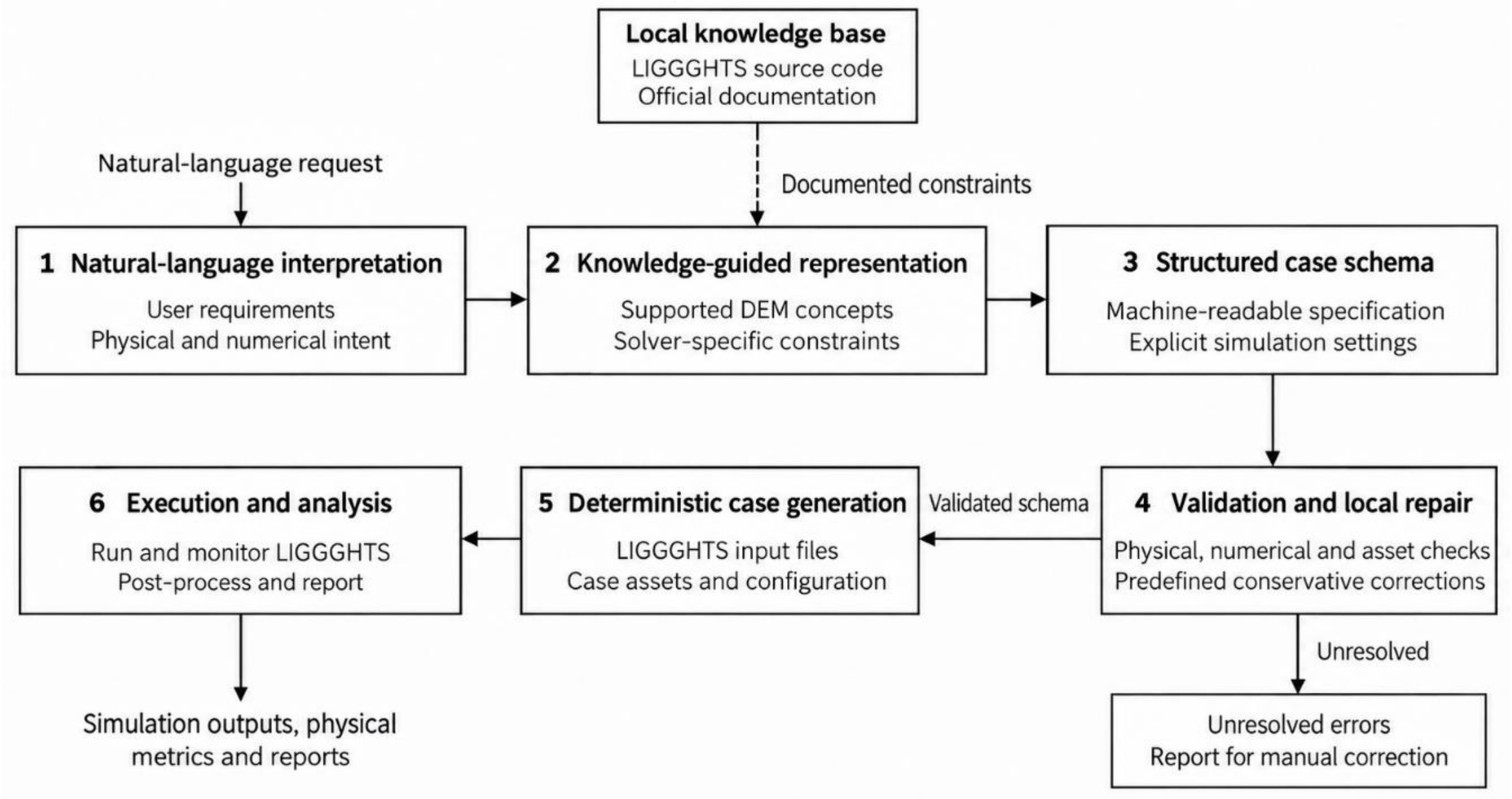

Fig. 1 Overall workflow of the proposed GranularGPT framework.

The knowledge base, structured schema, physics-aware validation, and conservative repair mechanisms therefore serve as supporting components of the end-to-end workflow. Together, they constrain the LLM-assisted generation process to supported solver capabilities and reduce unsupported commands, inconsistent parameters, and invalid simulation configurations. This architecture enables GranularGPT to connect natural-language interaction with the complete DEM workflow rather than treating input-script generation as an isolated task.

A key design principle is the separation of natural-language interpretation from solver execution. Rather than directly generating executable LIGGGHTS commands, the LLM first produces a structured representation of the intended simulation. This intermediate representation provides a common interface between the language model and the DEM solver, allowing the simulation specification to be inspected, validated, modified, and regenerated before execution. The same representation is retained throughout the workflow to support reproducible case generation, restart operations, parameter variations, and downstream analysis.

### 2.1.1 Natural-language interpretation

The first component of GranularGPT converts a natural-language description of a DEM problem into an initial structured representation of the intended simulation. The LLM interprets the user's requirements and identifies the physical, geometric, numerical, computational, and analysis information required to construct the simulation. Instead of directly generating executable LIGGGHTS commands, the model maps the user description to predefined simulation concepts and schema fields, which are subsequently processed by the knowledge-guided representation and validation modules.

The natural-language input is interpreted across five principal categories: (i) physical properties, including particle size, density, Young's modulus, Poisson's ratio, coefficient of restitution, coefficient of friction, and rolling-friction coefficient; (ii) simulation geometry, including simulation regions, walls, STL meshes, and generated geometries; (iii) simulation configuration, including contact models, timestep, runtime, neighbour settings, restart requirements, and parallel-execution options; (iv) output specifications, including dump formats, monitoring requirements, and requested output quantities; and (v) analysis requirements, including particle statistics, velocity distributions, contact-force characteristics, force-chain analysis, discharge behaviour, mixing, compaction, thermal quantities, and other DEM-related metrics. This interpretation step establishes the initial simulation intent that is subsequently transformed into an executable and analysable DEM case.

### 2.1.2 Knowledge-guided DEM representation

A natural-language description does not necessarily contain all solver-specific information required to construct a valid DEM simulation. GranularGPT therefore incorporates a local knowledge base derived from the LIGGGHTS source code and official documentation. The knowledge base provides solver-specific information needed to translate high-level simulation requirements into supported DEM configurations. The knowledge base contains catalogues of supported material models, contact laws, constitutive parameters, unit systems, particle-insertion methods, boundary conditions, geometric objects, mesh walls, neighbour configurations, restart operations, and output options. During case construction, these catalogues are used to associate extracted user requirements with supported LIGGGHTS capabilities and to identify required parameters and dependencies.

The knowledge-guided representation serves two main purposes. First, it improves the traceability of generated simulation components by associating them with documented solver functionality. Second, it constrains the generation process to supported configurations, thereby reducing the likelihood of unsupported commands, incompatible model combinations, and incomplete parameter definitions. The knowledge base therefore acts as a domain-specific interface between the LLM and the DEM solver rather than as an independent source of simulation decisions.

### 2.1.3 Structured DEM case schema

As shown in Fig. 2, the structured DEM case schema serves as an intermediate representation that enables communication and coordination among the different components of GranularGPT. Rather than directly generating executable input scripts, the framework first converts the interpreted user requirements into a machine-readable specification of the intended simulation. The schema records the information required to define and reproduce a DEM case, including simulation metadata, unit systems, atom styles, material properties, contact models, particle templates, particle size distributions, simulation regions, boundary conditions, gravity, wall definitions, STL meshes, particle insertion strategies, runtime controls, restart configurations, output requests, and required external assets. Each field is associated with predefined constraints that are checked before executable input generation.

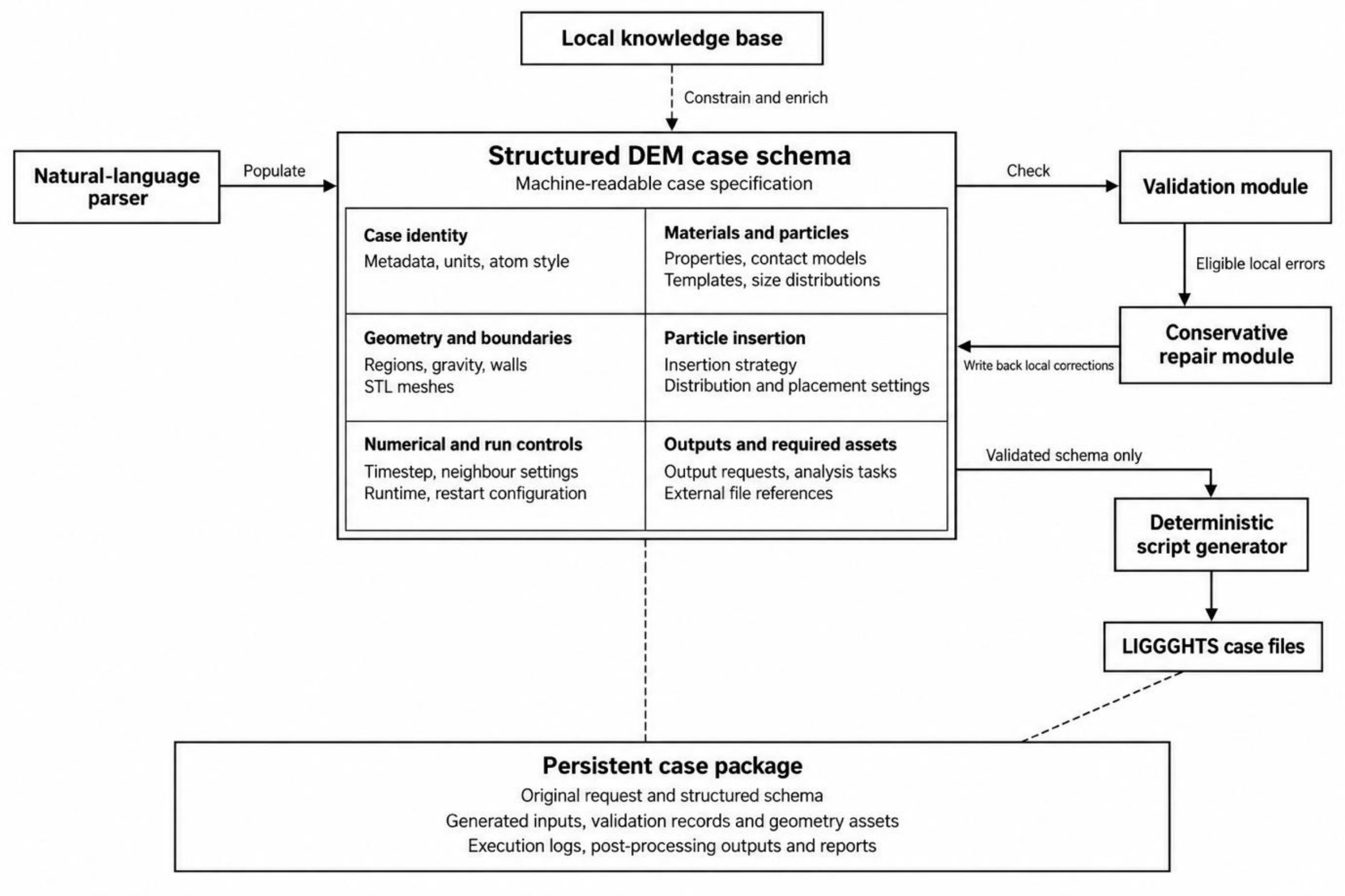

Fig. 2 Intermediate DEM case schema interfaces in GranularGPT.

Separating simulation intent from executable syntax allows the complete simulation specification to be inspected independently of solver commands. The representation can also be reused to generate parameter variations, reconstruct restart cases, and compare different simulation configurations. For each simulation, GranularGPT maintains a persistent case directory containing the original natural-language request, structured schema, generated LIGGGHTS input files, validation information, runtime logs, post-processing outputs, reports, and required geometry or other external assets. This persistent representation supports inspection, regeneration, rerunning, and comparison of simulation cases.

### 2.1.4 Physics-aware validation and conservative repair

Before execution, each structured case undergoes automated validation to assess its physical consistency, numerical stability, software compatibility, and required external resources. Validation is performed on the structured case representation before executable LIGGGHTS commands are generated, allowing inconsistencies to be identified at the simulation-definition level.

Physical validation evaluates the completeness and compatibility of material properties, contact models, particle templates, insertion strategies, wall definitions, restart configurations, and output requests. Numerical validation evaluates timestep stability using the Rayleigh timestep criterion, checks neighbour settings, evaluates particle-insertion feasibility, and verifies relevant runtime parameters. Resource validation checks the availability and consistency of referenced STL meshes, restart files, particle templates, and other required external assets.

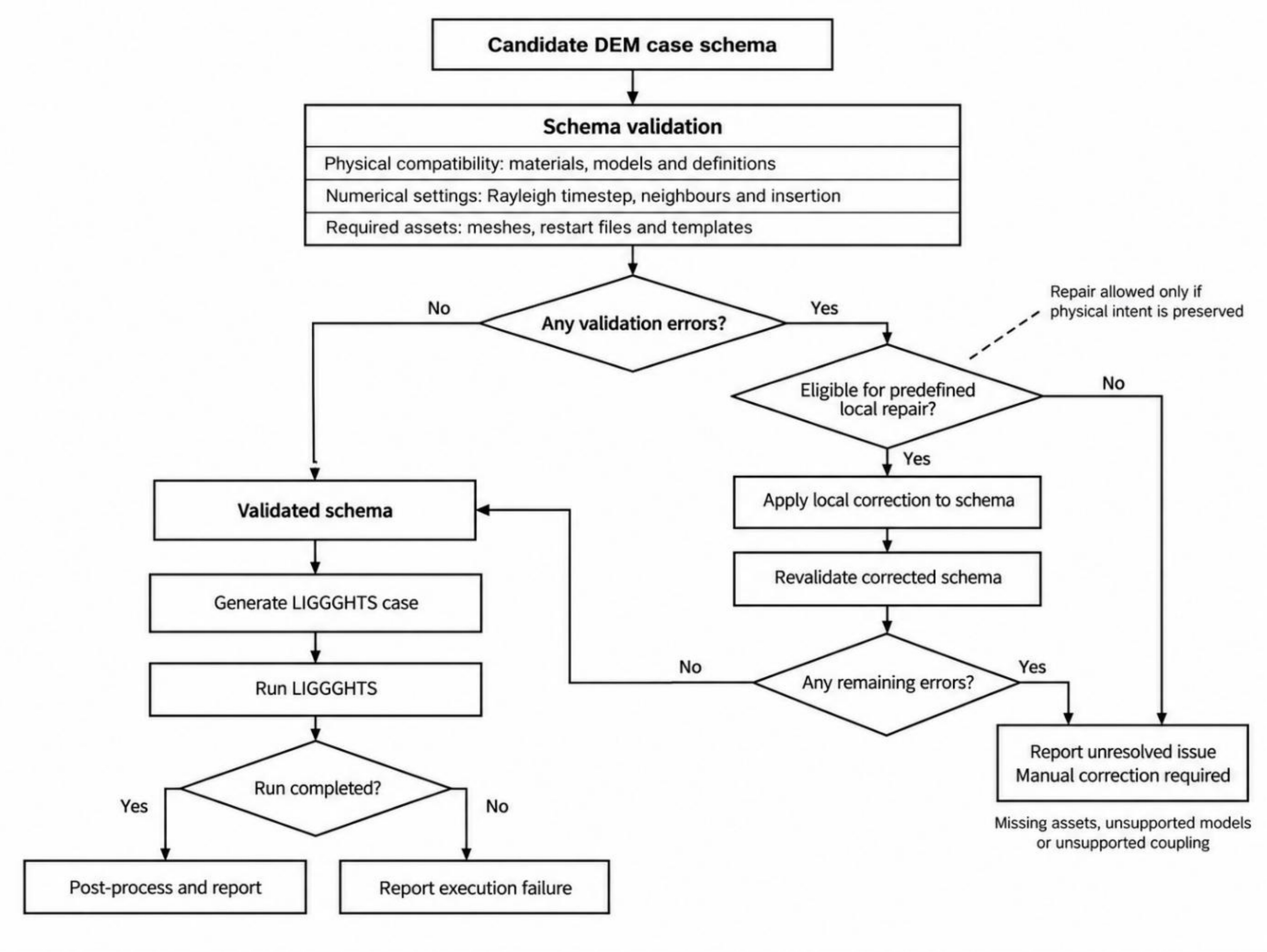

Fig. 3 Schema validation and conservative local repair in GranularGPT.

When an isolated inconsistency can be corrected without changing the intended physical model, GranularGPT applies a predefined conservative repair. Examples include restoring a valid unit system, replacing non-positive timestep values, normalising particle distributions, reducing an excessive timestep, adjusting neighbour parameters, correcting infeasible particle-insertion conditions, and disabling unsupported output requests. Repairs are restricted to predefined local transformations and are not used to infer scientifically significant user specifications. For example, missing geometries, unsupported contact laws, or other conditions requiring a substantive physical assumption are not automatically invented by the framework. Such cases are retained as unresolved errors and explicitly reported to the user. This conservative strategy is intended to improve workflow robustness while avoiding silent modification of scientifically meaningful simulation assumptions.

**2.1.5 Deterministic simulation generation**

Following successful validation, the structured DEM case is deterministically translated into a complete LIGGGHTS simulation. The generated input files contain the simulation components required by the validated case, including unit definitions, atom styles, simulation domains, neighbour settings, material-property definitions, contact models, gravity, particle templates, insertion mechanisms, wall and mesh definitions, integration commands, output configurations, restart operations, and runtime controls.

For newly constructed simulations, the framework generates the simulation domain, defines particle interactions and boundary conditions, inserts particles according to the specified strategy, and executes the requested simulation. For continuation workflows, existing restart files can be loaded and the required simulation environment reconstructed without regenerating the particle assembly. This enables restart-based workflows and parameter variations to be managed within the same case representation.

The deterministic generator provides a reproducible transition from the validated simulation specification to solver input. Under the same schema, software version, and configuration environment, the same input case is generated consistently. Each generated simulation is retained as part of the corresponding persistent case package together with its configuration, execution information, and outputs.

#### 2.1.6 Simulation execution and automated analysis

The generated simulation cases are executed through LIGGGHTS using either serial or MPI-based parallel execution. During execution, GranularGPT records solver status, runtime information, monitoring outputs, and error logs. Long-running simulations can be executed in detached mode while retaining their execution information within the corresponding case directory. The workflow also supports restart operations and parameter variations, allowing related simulation cases to be generated and managed from a common structured representation.

Following simulation completion, the post-processing module identifies the available simulation outputs and automatically extracts relevant physical and process-level quantities. Standard analyses include particle count, particle size distribution, spatial particle distribution, centroid position, translational and angular velocity statistics, contact force characteristics, and granular temperature indicators. Depending on the simulation objective, case-specific analyses can additionally include discharge rate, mixing index, force chain characteristics, compaction response, and thermal field quantities. The extracted quantities are converted into standardized quantitative reports that can be used to inspect individual simulations, compare parameter variations, and evaluate the resulting DEM behaviour. Thus, the workflow extends beyond simulation generation and execution to include automated interpretation of the resulting numerical data.

### 2.2 DEM formulation

GranularGPT employs the established DEM to simulate the dynamic behaviour of granular materials. The framework does not introduce a new DEM formulation; instead, it provides an LLM-assisted workflow for constructing, executing, and analysing DEM simulations using formulations implemented in LIGGGHTS. The translational and rotational motions of individual particles are obtained by solving Newton's equations of motion:

$$m_i \frac{d\boldsymbol{v}_i}{dt} = \sum_j \left(\boldsymbol{f}_{ij}^{n} + \boldsymbol{f}_{ij}^{t}\right) + m_i \boldsymbol{g} \tag{1}$$

$$I_i \frac{d\boldsymbol{\omega}_i}{dt} = \sum_j \left(\boldsymbol{T}_{ij}^{t} + \boldsymbol{T}_{ij}^{r}\right) \tag{2}$$

$$m_i c_{p,i} \frac{dT_i}{dt} = \sum_j Q_{ij} + Q_{rad} + Q_{source} \tag{3}$$

where $m_i$, $\boldsymbol{v}_i$, $I_i$, $\boldsymbol{\omega}_i$, $c_{p,i}$ and $T_i$ denote the mass, translational velocity, moment of inertia, angular velocity, specific heat capacity, and temperature for a particle $i$, respectively. $\boldsymbol{f}_{ij}^{n}$ and $\boldsymbol{f}_{ij}^{t}$ represent the normal and tangential interaction forces between particles $i$ and $j$, which consist of the contact force $\boldsymbol{f}_{c,ij}^{n}/\boldsymbol{f}_{c,ij}^{t}$ and damping force $\boldsymbol{f}_{d,ij}^{n}/\boldsymbol{f}_{d,ij}^{t}$, i.e., $\boldsymbol{f}_{ij}^{n} = \boldsymbol{f}_{c,ij}^{n} + \boldsymbol{f}_{d,ij}^{n}$ and $\boldsymbol{f}_{ij}^{t} = \boldsymbol{f}_{c,ij}^{t} + \boldsymbol{f}_{d,ij}^{t}$. The torques terms $\boldsymbol{T}_{ij}^{t}$ and $\boldsymbol{T}_{ij}^{r}$ arise from tangential forces and rolling resistance. $Q_{ij}$, $Q_{rad}$, and $Q_{source}$ represent particle-particle heat conduction, thermal radiation and heat source, respectively.

GranularGPT supports multiple contact formulations available in LIGGGHTS. Unless otherwise specified, the demonstration cases in this study employ the Hertz-Mindlin contact formulation. The selected contact formulation determines the material and interaction parameters required for a valid simulation, including Young's modulus, Poisson's ratio, coefficient of restitution, coefficient of friction, and rolling-friction coefficient. These dependencies are represented in the framework knowledge base and checked during case validation before input-file generation. Consequently, the physical requirements of the selected DEM formulation are explicitly connected to the structured case representation and the subsequent simulation-generation process.

### 2.3 Simulation setup and implementation

GranularGPT employs GPT-5.5 as its LLM agent to interpret natural-language requests and construct structured DEM case specifications. Given a user request, GranularGPT identifies the required physical properties, geometry, simulation parameters, boundary conditions, output requirements, and analysis objectives, and maps them to the structured DEM case schema described in Section 2.1. The framework is executed on a Linux workstation running Ubuntu 18.04.2 LTS, equipped with an Intel Core i7-8850H CPU and 32 GB of RAM. LIGGGHTS is executed using the same computational environment for all cases unless otherwise specified. A demonstration of the overall workflow is provided in Movie S1 of the Supplementary Materials.

# 3. Results and discussion

## 3.1 End-to-end demonstration from a single natural-language request

Fig. 4 demonstrates the complete GranularGPT workflow initiated from a single natural-language request. In this example, the user requests a DEM simulation of hopper discharge using LIGGGHTS, including hopper geometry generation, particle insertion, gravitational settling, outlet opening, discharge simulation, and post-processing. As shown in Fig. 4(a), the request specifies the intended simulation procedure, including a two-stage filling and discharge workflow, hopper dimensions, and required output files. Fig. 4(b) shows the corresponding response generated by GranularGPT. Based on the natural-language request, the framework constructs the simulation case, identifies the required input scripts and geometry files, executes the simulation, and summarizes the resulting discharge behaviour. The generated case includes the hopper STL geometry, LIGGGHTS input scripts for the filling and discharge stages, geometry summaries, simulation outputs, and post-processing files.

The resulting quantitative outputs include particle count, discharged mass, discharge completion time, peak discharge rate, and the final number of particles remaining in the hopper. This example demonstrates that the framework is not limited to translating natural-language descriptions into solver input files. Instead, a single user request initiates a sequence of operations spanning simulation construction, execution, output extraction, and quantitative analysis.

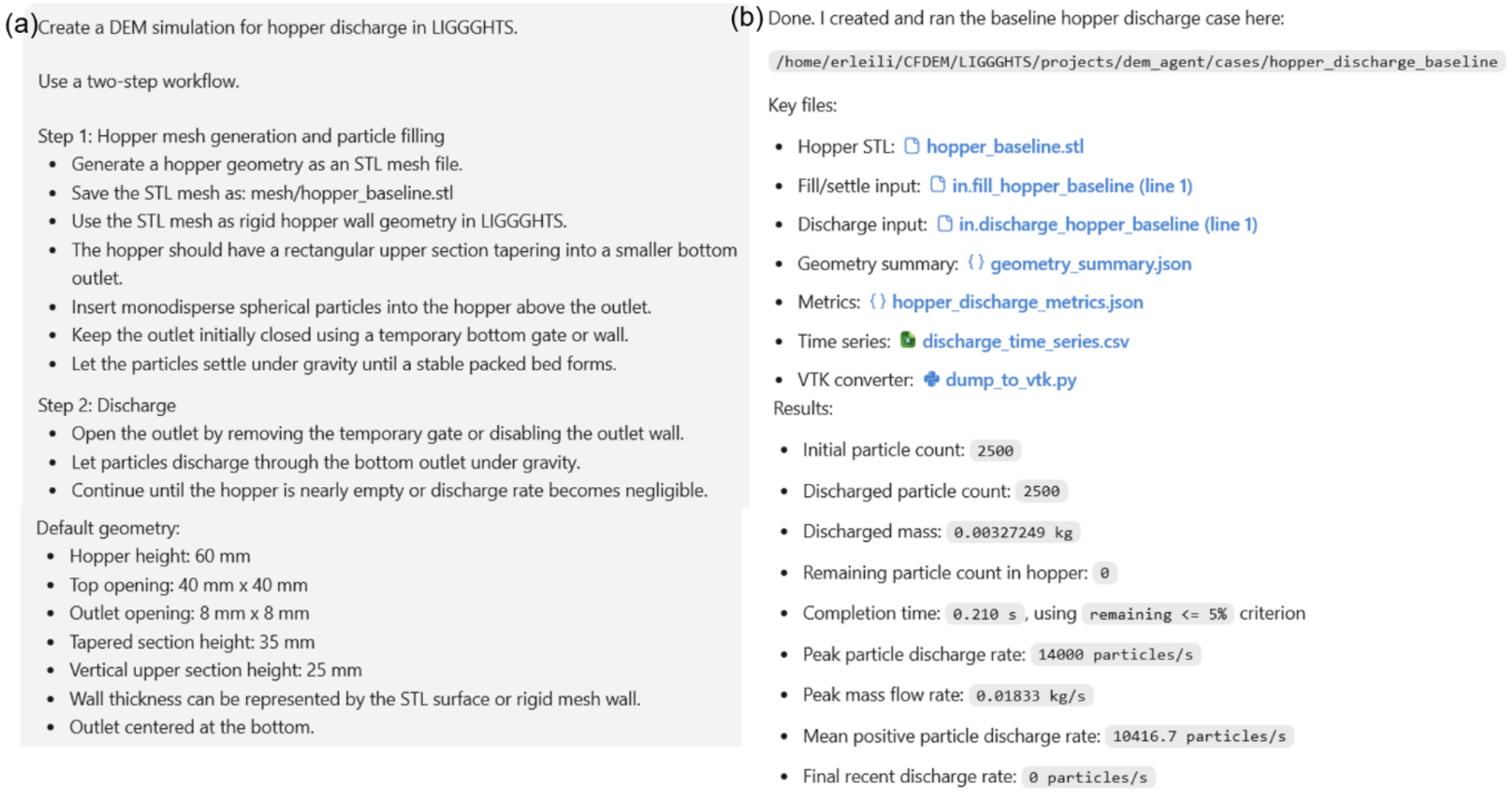

Fig. 4. End-to-end demonstration of GranularGPT: (a) natural-language request for hopper discharge simulation and (b) corresponding generated response.

Fig. 5 presents the generated hopper-discharge case and representative simulation results. Fig. 5(a) shows the resulting case directory, including input scripts, mesh files, post-processing scripts, and simulation outputs. Fig. 5(b) shows the generated hopper geometry, consisting of a rectangular upper section connected to a tapered lower section and a bottom outlet. Fig. 5(c) illustrates the particle-filling stage, in which spherical particles are inserted into the hopper and settle under gravity to form a packed bed. Fig. 5(d) shows the subsequent discharge stage after the outlet is opened, with particles flowing through the bottom outlet and accumulating on the bottom wall. The particle visualizations provide a qualitative indication of the simulated flow behaviour, while the associated numerical outputs provide quantitative measures of the discharge process. Together, Figs. 4 and 5 demonstrate the complete transition from a high-level engineering request to a runnable DEM case and quantitative simulation results.

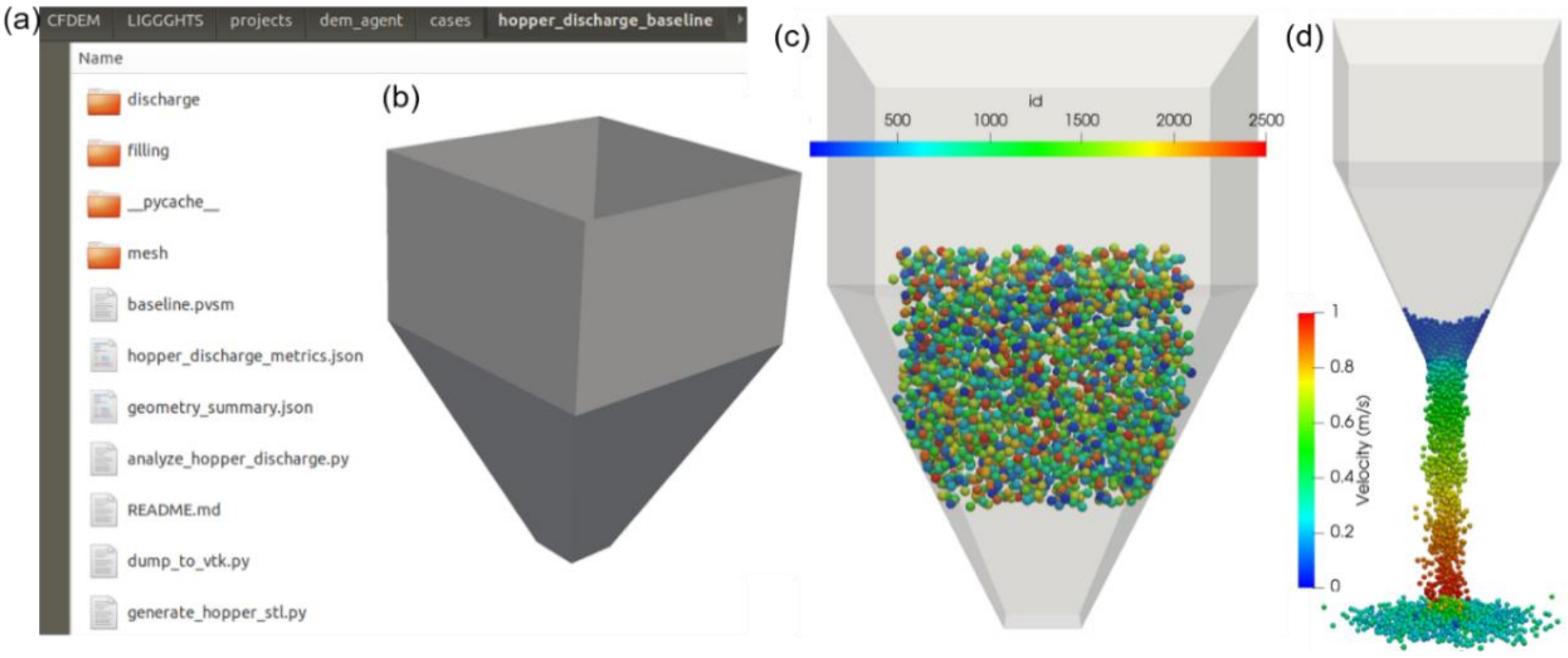

Fig. 5. GranularGPT-generated DEM case and simulation results: (a) case structure and generated files, (b) hopper geometry, (c) particle filling and gravitational settling, and (d) particle discharge through the outlet.

### 3.2 Evaluation across DEM scenarios

To evaluate whether the end-to-end workflow generalizes beyond a single demonstration case, 60 normal-condition cases are constructed and evaluated. The cases comprise 12 basic DEM scenarios and 48 parameter-variation cases, with four additional variants associated with each basic scenario. The scenarios cover a range of granular processes and modelling configurations, including particle packing, chute flow, powder spreading, hopper discharge, rotating drum, screw conveyor, blade mixing, conveyor transport, cohesive powder flow, multi-sphere clumps, compaction, and heat transfer. The case groups are summarized in Table 1, while the parameter variation settings are provided in Table 2.

A simulation case is considered successful when the case files are generated, LIGGGHTS is launched successfully, the prescribed final timestep is reached, and no fatal solver error occurs. This definition evaluates the workflow at the level of complete simulation execution rather than only assessing whether an input script is syntactically generated. Of the 60 evaluated cases, 55 completed successfully and 5 failed, corresponding to an overall success rate of 91.7%. Importantly, the five failures were concentrated in the screw-conveyor scenario and occurred across its five parameter variants. The remaining scenarios and their associated parameter variations completed successfully.

Table 1. Basic implemented scenarios and parameter variation cases

| Case group | Number | Purpose |
|---|---|---|
| Basic implemented scenarios | 12 | Evaluation of normal condition end-to-end DEM workflow generation |
| Parameter variation cases | 48 | Evaluation of workflow robustness to variations in material, contact, geometry, output, and runtime settings |
| Total | 60 | Normal condition evaluation |

Table 2. Parameter variation case settings

| Variant | Modification | Purpose |
|---|---|---|
| Case 1 | Baseline | Evaluation of the default case |
| Case 2 | Particle material variation | Evaluation of material property parsing |
| Case 3 | Contact variation | Evaluation of contact model setup |
| Case 4 | Geometry/boundary variation | Evaluation of domain and wall handling |
| Case 5 | Output/runtime variation | Evaluation of post-processing and runtime controls |

Fig. 6 presents representative results from the evaluated cases. The examples span particle packing, gravity-driven chute flow, powder spreading, rotating-drum motion, cohesive powder flow, multi-sphere clumps, blade mixing, conveyor transport, compaction, and heat transfer. These cases demonstrate that GranularGPT can construct and execute simulations involving different particle representations, boundary conditions, motion modes, and physical processes within a common workflow.

The five unsuccessful cases provide an informative indication of the current limitations of the framework. Their failure was traced to the automatically generated screw-conveyor STL geometry. The generated mesh contained poor-quality shaft triangles and extremely thin elements, causing LIGGGHTS to terminate during mesh initialization because the geometry did not satisfy the required mesh and curvature-tolerance conditions. GranularGPT attempted automatic correction for these cases, but the resulting geometries remained incompatible with the solver. Integration with dedicated geometry-generation and mesh-quality-validation tools could improve the robustness of such cases in future versions.

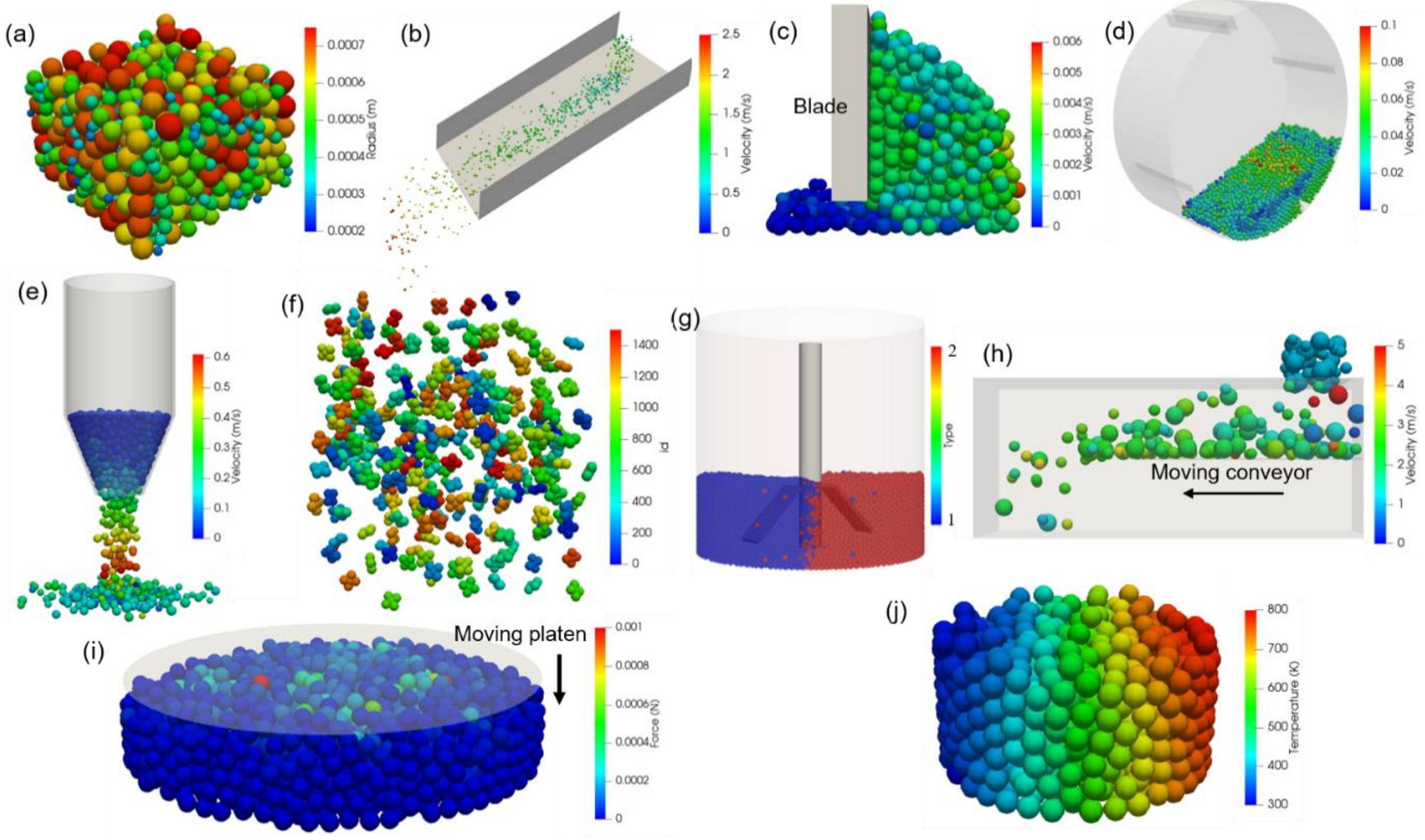

Fig. 6 Representative simulation results generated by GranularGPT: (a) particle packing, (b) chute flow, (c) powder spreading, (d) rotating drum, (e) cohesive powder flow, (f) multi-sphere clumps, (g) blade mixer, (h) conveyor transport, (i) compaction, and (j) heat transfer.

### 3.3 Error detection and conservative repair

In addition to the normal condition evaluation, 15 intentionally corrupted cases are constructed to assess the validation and error-handling capabilities of GranularGPT. The cases are divided into 10 repairable errors and 5 non-repairable conditions. The evaluation is designed to test whether the framework could distinguish between local specification errors that can be corrected safely and conditions that require additional user information or capabilities outside the implemented workflow. A demonstration of the error detection and repair is provided in Movie S2 of the Supplementary Materials.

The repairable cases include unsafe timestep, non-positive timestep, excessive target volume fraction, negative particle radius, zero particle density, negative Young's modulus, invalid Poisson's ratio, invalid output frequency, zero simulation steps, and an invalid boundary condition. For these cases, successful handling requires the framework to identify the invalid specification, apply a predefined conservative correction where appropriate, regenerate the simulation case, and produce a runnable configuration. All 10 repairable errors are detected. Nine cases are successfully repaired and subsequently become runnable, corresponding to a 90% repair success rate. The remaining case involves an excessively high target volume fraction of 0.9. GranularGPT reduces the specified value to 0.6 according to its predefined repair rule; however, the corrected case still fails to launch successfully.

The five non-repairable cases involve missing external assets, unavailable geometry, unsupported models, or capabilities outside the current workflow. These include a missing manually supplied STL wall, a missing restart file, an unsupported or unknown contact model, a missing multi-sphere template file, and a gas-solid fluidization request requiring CFD-DEM coupling. For these cases, the expected behaviour is explicit identification and reporting rather than automatic modification. All five non-repairable cases are correctly identified and reported. The resulting performance is summarized in Table 3.

Table 3. Error-handling performance

| Metric | Result |
|---|---|
| Repairable errors detected | 10/10 (100%) |
| Repairable errors successfully resolved | 9/10 (90%) |
| Non-repairable errors correctly reported | 5/5 (100%) |

The error-handling results demonstrate that the validation and repair mechanisms can detect a range of invalid DEM specifications while resolving most local errors without requiring manual intervention. More importantly, the framework distinguishes between repairable configuration errors and non-repairable scientific or resource-level deficiencies. Missing assets and unsupported capabilities are not silently replaced by inferred assumptions. This conservative behaviour is important for scientific simulation because automatically introducing an unsupported physical model or inventing missing geometry could produce apparently valid but scientifically misleading results.

### 3.4 Automated physical metrics and post-processing

The end-to-end workflow extends beyond simulation construction and execution to automated extraction and analysis of DEM results. GranularGPT identifies the output files generated during simulation, including log files, particle data files, mesh files, and monitoring outputs, and converts these data into quantitative summaries and visualization formats. Where applicable, particle and mesh data can also be converted into ParaView-compatible formats for subsequent visualization. A demonstration of the automated post-processing is provided in Movie S3 of the Supplementary Materials.

For each available particle output frame, the post-processing module calculates general DEM quantities including particle count, particle type distribution, particle size statistics, particle positions, velocity statistics, contact force statistics, angular velocity statistics, particle bounds, and a granular temperature proxy. These general quantities provide a standardized description of the simulated state. In addition, the framework supports case-specific analyses such as particle size distributions, bulk density measures, layer height maps, velocity profiles, discharge rate curves, mixing index time series, force chain visualizations, compaction force-displacement curves, and thermal field metrics.

Fig. 7 presents representative particle size and bulk state analyses. The particle size distribution gives D10 of 25.620 μm, D50 of 32.092 μm, and D90 of 41.103 μm, showing that most particles are concentrated in the approximate range of 25-41 μm. The peak probability occurs near 32-34 μm, while the cumulative distribution approaches 100% at about 50-55 μm (Fig. 7a). For the compaction case, the framework calculates the solid volume fraction as a measure of bed density. The packing fraction increases from approximately 41% at the initial state to approximately 57% after compaction, with most of the densification occurring between 0.84 and 0.92 s before reaching a relatively stable value (Fig. 7b).

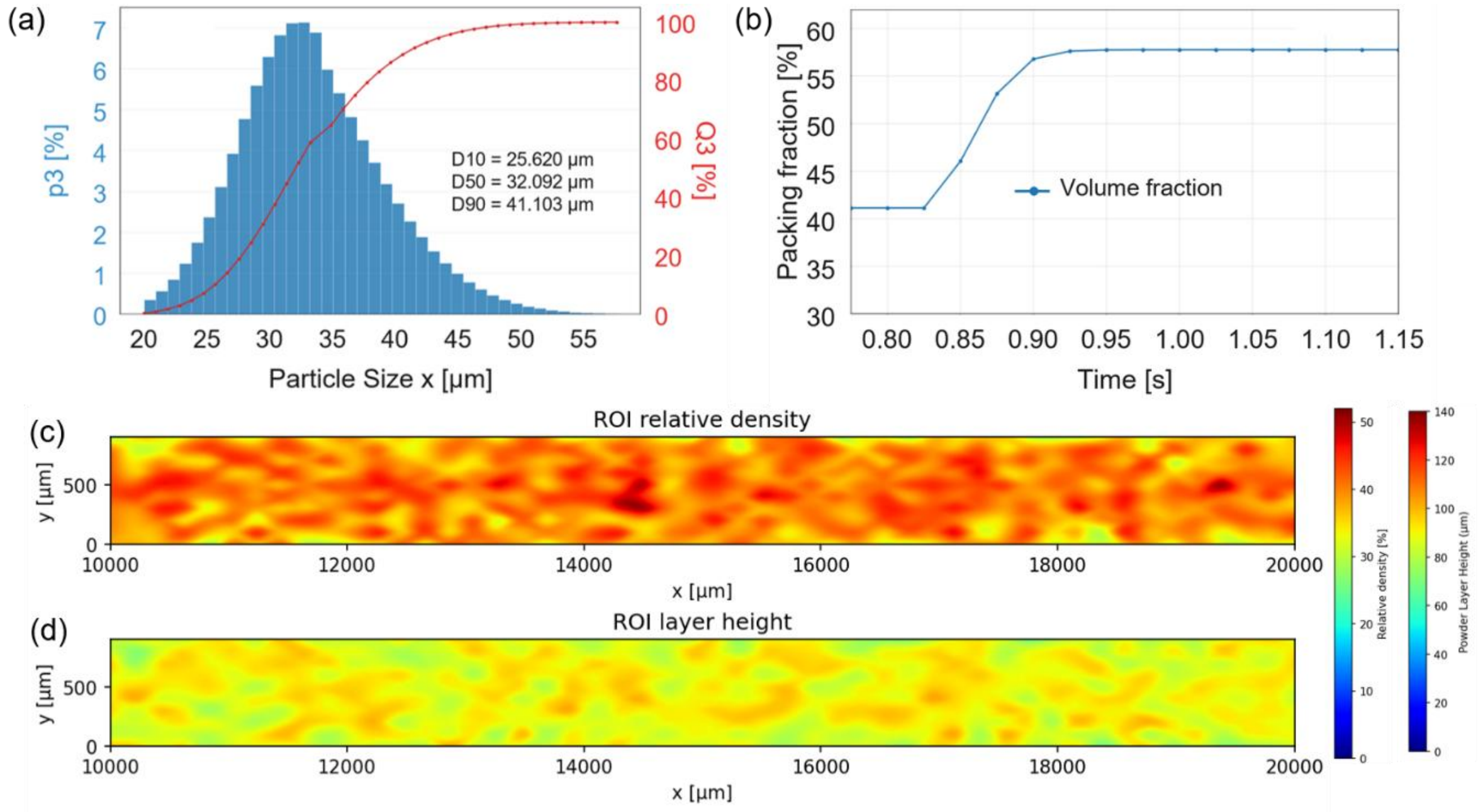

Fig. 7 Physical metrics generated by GranularGPT: (a) particle size distribution, (b) packing fraction evolution during compaction, (c) relative packing density map, and (d) layer height map after powder spreading.

For powder spreading, GranularGPT generates spatial maps of relative density and layer height within the specified region of interest (Figs. 7c and d). The relative density map indicates spatially heterogeneous packing, with most regions exhibiting intermediate relative densities of approximately 30-45% and localized regions approaching 50%. The layer height map shows a continuous deposited powder layer with spatial variations in thickness. Most of the layer is approximately 80-110 μm thick, with local regions ranging from approximately 60 to 140 μm. These outputs demonstrate that the post-processing module can convert particle-level simulation data into spatially resolved process-level quantities.

Fig. 8 demonstrates post-processing of dynamic granular flow and mixing cases. The rotating drum visualization shows particle velocity within the drum, providing a qualitative representation of the flow regime (Fig. 8a). The rotating drum snapshot shows particle velocity up to approximately 0.066 m/s. Higher velocities appear near the flowing surface, while lower velocities dominate inside the bulk bed. For hopper discharge, the framework calculates the outlet mass flow rate and cumulative discharged mass, providing quantitative measures of both instantaneous discharge behaviour and total material discharge (Figs. 8b and c). The hopper discharge mass flow rate starts after outlet opening at approximately 0.10 s. During discharge, the mass flow rate fluctuates mainly between 0.02 and 0.04 kg/s, with a peak near 0.045 kg/s. The cumulative discharged mass increases nearly linearly after 0.10 s, reaching approximately 0.014 kg at the end of the simulation.

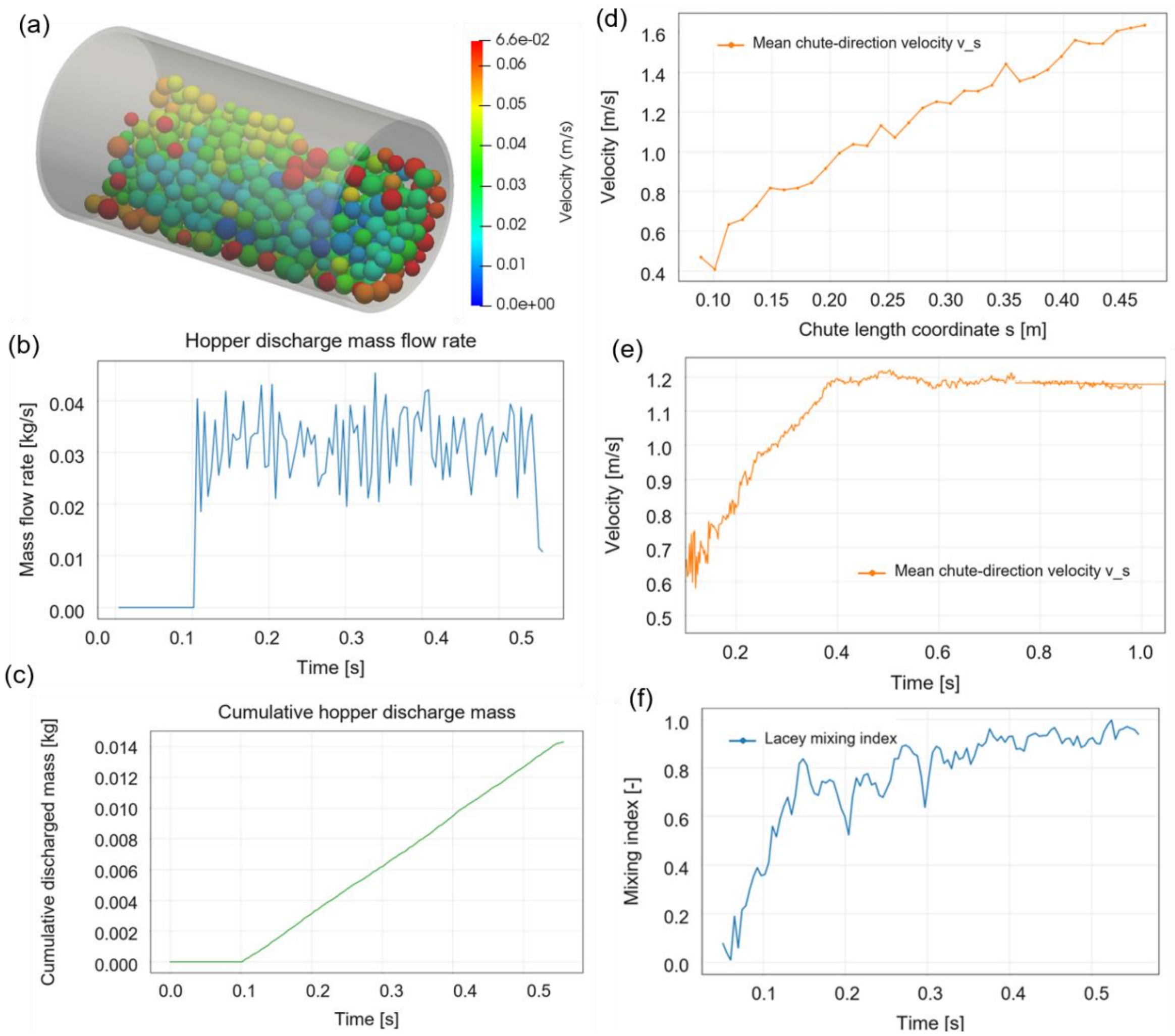

Fig. 8 Flow, discharge, and mixing metrics generated by GranularGPT: (a) particle velocity snapshot in a rotating drum, (b) hopper discharge mass flow rate, (c) cumulative discharged mass, (d) chute direction velocity profile along the inclined chute coordinate at a specified time, (e) temporal evolution of the mean chute direction particle velocity, and (f) Lacey mixing index for binary drum mixing.

For chute flow analysis, the framework extracts the mean particle velocity in the chute direction as a function of the inclined coordinate, enabling quantification of streamwise flow development at a selected time (Fig. 8d). The chute-flow velocity increases along the inclined direction from approximately 0.4-0.5 m/s upstream to about 1.6 m/s downstream. This trend indicates sustained particle acceleration along the chute, with small local fluctuations caused by granular interactions. The temporal evolution of the mean chute-direction velocity is also calculated, as shown in Fig. 8e. The mean chute-direction velocity rises rapidly from about 0.6 m/s to approximately 1.2 m/s within the first 0.4 s. After that, the velocity remains close to 1.17-1.22 m/s, suggesting a quasi-steady transport regime. For binary mixing cases, the Lacey mixing index is calculated as a function of time (Fig. 8f). The Lacey mixing index increases from near 0 to above 0.8 during the early mixing stage. It later fluctuates around 0.9-1.0, indicating that the particle system approaches a well-mixed state.

Fig. 9 presents examples of force chain, compaction force, and thermal post-processing. Force chain analysis identifies strong particle-particle and particle-wall contacts and exports the corresponding contact network for visualization (Fig. 9a), enabling examination of load-bearing structures within the granular assembly. The force-chain visualization shows a heterogeneous contact network, where most contacts carry relatively weak forces. In the compaction case, the compressive force acting on the platen is plotted against platen displacement (Fig. 9b), providing a macroscopic force-displacement response of the particle bed. The compaction force remains close to zero at small displacement and then increases rapidly after approximately 1.5 mm. Near the platen-stop position, the normal force reaches about 50 N, indicating strong resistance from the densified particle bed. For thermal DEM cases, the framework extracts the mean particle temperature and heating rate as functions of time (Fig. 9c). The mean particle temperature increases from approximately 300 K to about 415-420 K during heating. Meanwhile, the heating rate decreases from above 400 K/s to below 100 K/s, showing that the system heats rapidly at first and then gradually slows. The thermal gradient analysis further reports the specified spatial temperature gradient, as shown in Fig. 9d. The mean thermal gradient rises sharply from near 0 to above 5000 K/m during the early stage. It reaches a maximum around 1.1-1.2 s and then slightly decreases to about 4800-5000 K/m near the end.

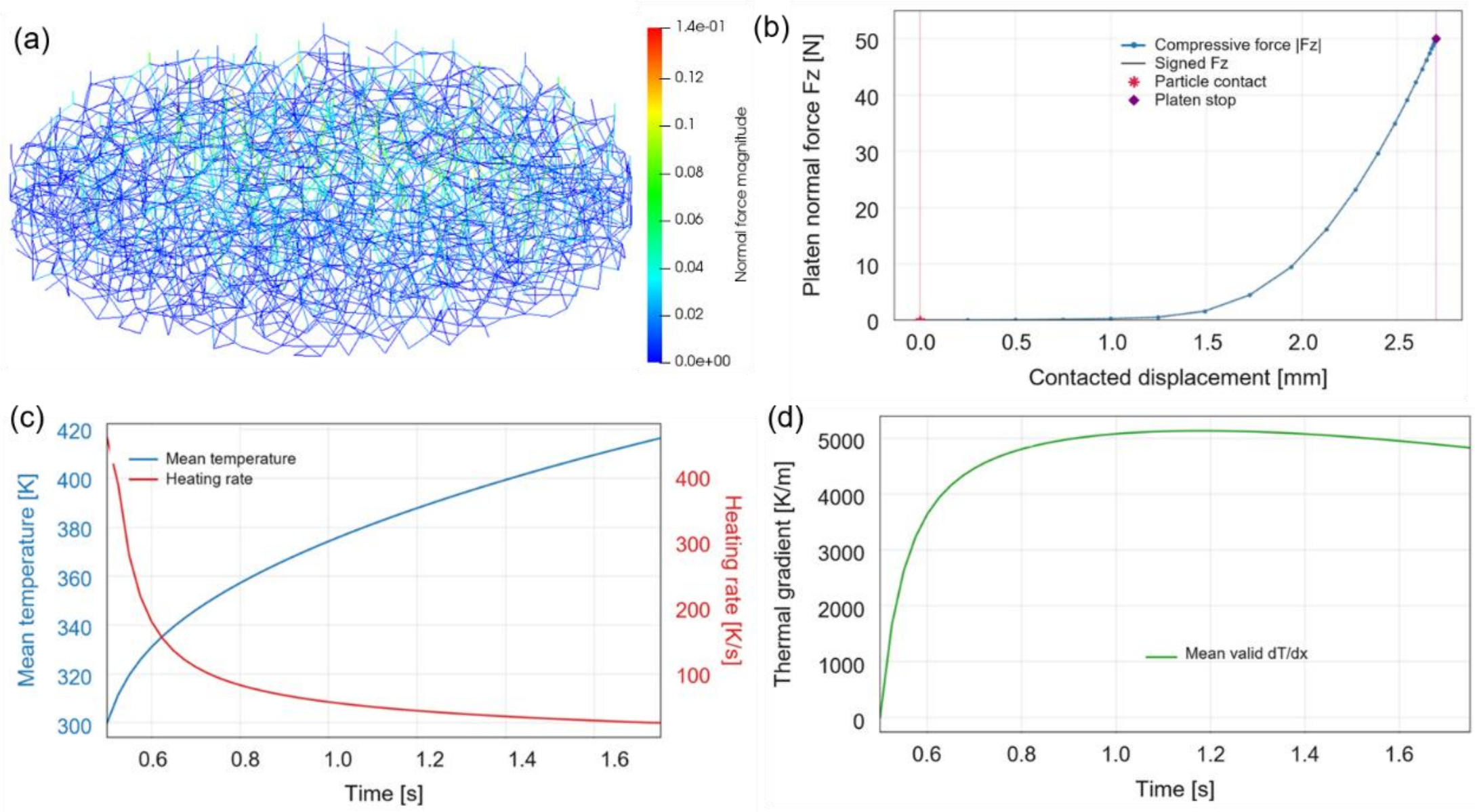

Fig. 9 Force chain, compaction, and thermal metrics generated by GranularGPT: (a) force-chain network visualization and (b) compressive platen force as a function of displacement, (c) particle temperature and heating rate, and (d) thermal gradient.

Taken together, Figs. 7-9 demonstrate that GranularGPT extends DEM automation beyond simulation generation and execution to quantitative interpretation of simulation outputs. The post-processing module can extract both general particle-level quantities and scenario-specific process metrics, allowing the same natural-language-driven workflow to produce simulation results and corresponding quantitative analyses. This capability is central to the end-to-end nature of the framework because the workflow terminates not at a successfully completed LIGGGHTS run, but at a set of physical quantities that can be used to evaluate and interpret the simulated process.

## 4. Conclusions

This study presented GranularGPT, an LLM-assisted agentic framework for end-to-end DEM simulation and automated analysis. The framework is developed to reduce the technical effort required to move from a natural-language description of a granular-process problem to an executable simulation and quantitative interpretation of the resulting data. Rather than limiting LLM use to input-script generation, GranularGPT integrates natural-language interpretation, structured DEM case construction, solver-specific knowledge, physics-aware validation, conservative error repair, deterministic simulation generation, execution management, and automated post-processing within a unified workflow.

The framework is evaluated using 60 normal-condition cases covering 12 DEM scenarios and associated parameter variations. Of these cases, 55 are completed successfully, corresponding to an overall workflow success rate of 91.7%. The evaluated scenarios include particle packing, chute flow, powder spreading, hopper discharge, rotating drums, screw conveyors, blade mixing, conveyor transport, cohesive powder flow, multi-sphere clumps, compaction, and heat transfer. These results demonstrate the feasibility of applying the same natural-language-driven workflow to a range of granular systems and simulation configurations.

The validation and error-handling capabilities are assessed using 15 intentionally corrupted cases. All 10 repairable errors are detected, and 9 are successfully resolved, giving a repair success rate of 90%. All five non-repairable cases are correctly identified and reported. These results highlight the importance of combining LLM-based case generation with structured representations and domain-specific validation.

GranularGPT also demonstrated automated extraction of a broad range of DEM metrics from simulation outputs. These included particle size distributions, particle counts, packing fractions, density and layer height maps, velocity profiles, discharge rates, mixing indices, force chain characteristics, compaction force-displacement responses, and thermal quantities. The integration of these capabilities enables the workflow to proceed beyond simulation execution to quantitative analysis and reporting, which is an important aspect of the proposed end-to-end approach.

Overall, GranularGPT demonstrates the potential of LLMs to support end-to-end DEM modelling, from simulation setup and execution to automated analysis. Its flexible workflow can be extended to a wide range of granular-material processes. Future work will focus on improving complex geometry generation, enabling multiphysics coupling, and developing more reliable and capable LLM-assisted DEM solvers.

## References

[1] C.S. Campbell, Granular material flows–an overview, Powder Technology 162(3) (2006) 208-229.
[2] L. Orefice, J. Khinast, DEM study of granular transport in partially filled horizontal screw conveyors, Powder Technology 305 (2017) 347-356.
[3] M. Houssais, C.P. Ortiz, D.J. Durian, D.J. Jerolmack, Onset of sediment transport is a continuous transition driven by fluid shear and granular creep, Nature communications 6(1) (2015) 6527.
[4] R. Vehring, Pharmaceutical particle engineering via spray drying, Pharmaceutical research 25(5) (2008) 999-1022.
[5] M.L. Hamad, K. Bowman, N. Smith, X. Sheng, K.R. Morris, Multi-scale pharmaceutical process understanding: From particle to powder to dosage form, Chemical engineering science 65(21) (2010) 5625-5638.
[6] R. Agrawal, Y. Naveen, Pharmaceutical processing–A review on wet granulation technology, International journal of pharmaceutical frontier research 1(1) (2011) 65-83.
[7] M. Li, A. Li, J. Zhang, Y. Huang, J. Li, Effects of particle sizes on compressive deformation and particle breakage of gangue used for coal mine goaf backfill, Powder Technology 360 (2020) 493-502.
[8] S. Liu, Q. Li, J. Song, Study on the grinding kinetics of copper tailing powder, Powder Technology 330 (2018) 105-113.
[9] B. Cuq, E. Rondet, J. Abecassis, Food powders engineering, between knowhow and science: Constraints, stakes and opportunities, Powder technology 208(2) (2011) 244-251.
[10] A. Raji, J. Favier, Model for the deformation in agricultural and food particulate materials under bulk compressive loading using discrete element method. I: Theory, model development and validation, Journal of food engineering 64(3) (2004) 359-371.
[11] E. Li, H.Y. Chia, Y. Ma, W. Yan, Mechanisms of fibre orientation and breakage in short fibre reinforced composites by fused filament fabrication, Composites Part B: Engineering 324 (2026) 113860.

[12] E. Li, W. Yan, Binder-powder interactions in binder jetting: Binder drying, layer shifting, and inter-layer binding, Additive Manufacturing 110 (2025) 104951.
[13] L. Wang, A. Yu, E. Li, H. Shen, Z. Zhou, Effects of spreader geometry on powder spreading process in powder bed additive manufacturing, Powder Technology 384 (2021) 211-222.
[14] H.H. Nersisyan, J.H. Lee, H.T. Son, C.W. Won, D.Y. Maeng, A new and effective chemical reduction method for preparation of nanosized silver powder and colloid dispersion, Materials Research Bulletin 38(6) (2003) 949-956.
[15] W. Songping, M. Shuyuan, Preparation of micron size copper powder with chemical reduction method, Materials Letters 60(20) (2006) 2438-2442.
[16] P.A. Cundall, O.D. Strack, A discrete numerical model for granular assemblies, geotechnique 29(1) (1979) 47-65.
[17] R.D. Mindlin, H. Deresiewicz, Elastic spheres in contact under varying oblique forces, Journal of Applied Mechanics 20 (1953) 327-344.
[18] H.P. Zhu, Z.Y. Zhou, R.Y. Yang, A.B. Yu, Discrete particle simulation of particulate systems: Theoretical developments, Chemical Engineering Science 62(13) (2007) 3378-3396.
[19] H.P. Zhu, Z.Y. Zhou, R.Y. Yang, A.B. Yu, Discrete particle simulation of particulate systems: A review of major applications and findings, Chemical Engineering Science 63(23) (2008) 5728-5770.
[20] Y. Cheng, S. Guo, H. Lai, Dynamic simulation of random packing of spherical particles, Powder Technology 107(1-2) (2000) 123-130.
[21] X.L. Deng, R.N. Davé, Dynamic simulation of particle packing influenced by size, aspect ratio and surface energy, Granular Matter 15(4) (2013) 401-415.
[22] M. Combarros, H. Feise, H. Zetzener, A. Kwade, Segregation of particulate solids: Experiments and DEM simulations, Particuology 12 (2014) 25-32.
[23] Y. Yu, H. Saxén, Experimental and DEM study of segregation of ternary size particles in a blast furnace top bunker model, Chemical engineering science 65(18) (2010) 5237-5250.
[24] S. Wu, M. Kou, J. Xu, X. Guo, K. Du, W. Shen, J. Sun, DEM simulation of particle size segregation behavior during charging into and discharging from a Paul-Wurth type hopper, Chemical engineering science 99 (2013) 314-323.
[25] F. Bertrand, L.-A. Leclaire, G. Levecque, DEM-based models for the mixing of granular materials, Chemical Engineering Science 60(8-9) (2005) 2517-2531.
[26] P. Liu, R. Yang, A. Yu, DEM study of the transverse mixing of wet particles in rotating drums, Chemical Engineering Science 86 (2013) 99-107.
[27] M. Alian, F. Ein-Mozaffari, S.R. Upreti, Analysis of the mixing of solid particles in a plowshare mixer via discrete element method (DEM), Powder Technology 274 (2015) 77-87.
[28] P.W. Cleary, DEM prediction of industrial and geophysical particle flows, Particuology 8(2) (2010) 106-118.
[29] S.K. Wilkinson, S.A. Turnbull, Z. Yan, E.H. Stitt, M. Marigo, A parametric evaluation of powder flowability using a Freeman rheometer through statistical and sensitivity analysis: A discrete element method (DEM) study, Computers & Chemical Engineering 97 (2017) 161-174.
[30] A. Anand, J.S. Curtis, C.R. Wassgren, B.C. Hancock, W.R. Ketterhagen, Predicting discharge dynamics from a rectangular hopper using the discrete element method (DEM), Chemical Engineering Science 63(24) (2008) 5821-5830.
[31] A. Anand, J.S. Curtis, C.R. Wassgren, B.C. Hancock, W.R. Ketterhagen, Predicting discharge dynamics of wet cohesive particles from a rectangular hopper using the discrete element method (DEM), Chemical Engineering Science 64(24) (2009) 5268-5275.
[32] P.W. Cleary, M.L. Sawley, DEM modelling of industrial granular flows: 3D case studies and the effect of particle shape on hopper discharge, Applied Mathematical Modelling 26(2) (2002) 89-111.
[33] C. Kloss, C. Goniva, A. Hager, S. Amberger, S. Pirker, Models, algorithms and validation for opensource DEM and CFD–DEM, Progress in Computational Fluid Dynamics, an International Journal 12(2-3) (2012) 140-152.
[34] C. Kloss, C. Goniva, LIGGGHTS–open source discrete element simulations of granular materials based on Lammps, Supplemental proceedings (2011) 781-788.
[35] L. Ouyang, J. Wu, X. Jiang, D. Almeida, C. Wainwright, P. Mishkin, C. Zhang, S. Agarwal, K. Slama, A. Ray, Training language models to follow instructions with human feedback, Advances in neural information processing systems 35 (2022) 27730-27744.

[36] J. Achiam, S. Adler, S. Agarwal, L. Ahmad, I. Akkaya, F.L. Aleman, D. Almeida, J. Altenschmidt, S. Altman, S. Anadkat, Gpt-4 technical report, arXiv preprint arXiv:2303.08774 (2023).
[37] D.A. Boiko, R. MacKnight, B. Kline, G. Gomes, Autonomous chemical research with large language models, Nature 624(7992) (2023) 570-578.
[38] A. M. Bran, S. Cox, O. Schilter, C. Baldassari, A.D. White, P. Schwaller, Augmenting large language models with chemistry tools, Nature machine intelligence 6(5) (2024) 525-535.
[39] T. Schick, J. Dwivedi-Yu, R. Dessì, R. Raileanu, M. Lomeli, E. Hambro, L. Zettlemoyer, N. Cancedda, T. Scialom, Toolformer: Language models can teach themselves to use tools, Advances in neural information processing systems 36 (2023) 68539-68551.
[40] T.D. Pham, A. Tanikanti, M. Keçeli, ChemGraph as an agentic framework for computational chemistry workflows, Communications Chemistry 9(1) (2026) 33.
[41] G. Liu, X. Ma, L. Zhang, J. Pascasio, J. Yang, Y. Chen, W.-L. Li, CatGo: Bridging CLI coding agents with interactive structure and workflow management for computational chemistry, Digital Discovery (2026).
[42] Z. Shi, C. Xin, T. Huo, Y. Jiang, B. Wu, X. Chen, W. Qin, X. Ma, G. Huang, Z. Wang, A fine-tuned large language model based molecular dynamics agent for code generation to obtain material thermodynamic parameters, Scientific Reports 15(1) (2025) 10295.
[43] S. Pandey, R. Xu, W. Wang, X. Chu, OpenFOAMGPT: A retrieval-augmented large language model (LLM) agent for OpenFOAM-based computational fluid dynamics, Physics of Fluids 37(3) (2025).
[44] L. Yue, N. Somasekharan, T. Zhang, Y. Cao, Z. Chen, S. Di, S. Pan, Foam-Agent: A large language model-based multi-agent framework for automating computational fluid dynamics workflows, Computer Methods in Applied Mechanics and Engineering 461 (2026) 119271.
[45] D. Park, H. Moon, S. Ryu, A self-correcting multi-agent LLM framework for language-based physics simulation and explanation, npj Artificial Intelligence 2(1) (2026) 10.
[46] S. Lv, L. Peng, W. Wu, Y. Yao, S. Jiao, W. Hu, Bridging language models and computational materials science: a prompt-driven framework for material property prediction, Materials Genome Engineering Advances 3(2) (2025) e70013.
[47] I. Takahara, T. Mizoguchi, B. Liu, Accelerated inorganic materials design with generative AI agents, Cell Reports Physical Science 6(12) (2025).